\documentclass[onecolumn,showpacs,preprintnumbers,amsmath,amssymb,aps,letterpaper]{revtex4}
\usepackage[toc,page,title,titletoc,header]{appendix}
\input{psfig.sty}

\begin{document}
\addtolength{\topmargin}{1.3cm}
%opening
\title{Cosmological constraints from Radial Baryon Acoustic Oscillation measurements and Observational Hubble data}
\author{Zhong-Xu Zhai$^1$}
\author{Hao-Yi Wan$^2$}
\author{Tong-Jie Zhang$^{3,4}$}
\email{tjzhang@bnu.edu.cn}
%\author{Jie Zhou$^2$}

%\address{Department of Astronomy, Beijing Normal University,
%Beijing, 100875, P.R.China}

\affiliation{$^1$Department of Physics, Beijing Normal University,
Beijing, 100875, P.R.China}
\affiliation{$^2$Beijing Planetarium, Beijing 100044, PR China}
\affiliation{$^3$Department of Astronomy, Beijing Normal University, Beijing, 100875, P.R.China}
\affiliation{$^4$Center for High Energy Physics, Peking University, Beijing 100871, P.R. China}
%\affiliation{$^5$Kavli Institute for Theoretical
%Physics China, CAS, Beijing 100190, P.R.China}

\begin{abstract}
We use the Radial Baryon Acoustic Oscillation (RBAO) measurements, distant type Ia supernovae
(SNe Ia), the observational $H(z)$ data (OHD) and the Cosmic Microwave Background (CMB) shift parameter
data to constrain cosmological parameters of $\Lambda$CDM and XCDM cosmologies and further examine the role of
OHD and SNe Ia data in cosmological constraints. We marginalize the likelihood function over $h$ by integrating the
probability density $P\propto e^{-\chi^{2}/2}$ to obtain the best fitting results and the confidence regions
in the $\Omega_{m}-\Omega_{\Lambda}$ plane. %The results show that the prior of $h=0.705\pm0.013$ can improve the constraints well.
With the combination analysis for both of the {\rm $\Lambda$}CDM and XCDM models,
we find that the confidence regions of 68.3\%, 95.4\% and 99.7\% levels using
OHD+RBAO+CMB data are in good agreement with that of SNe Ia+RBAO+CMB data which is consistent with the result of Lin et al's work.
With more data of OHD, we can
probably constrain the cosmological parameters using OHD data instead of SNe Ia data in the future.

\end{abstract}

%Uncomment for PACS numbers title message
\pacs{98.80.Es} \maketitle

\section{Introduction} \label{sec:introduction}
In modern cosmology, the discovery of the accelerating expansion of the universe is a great
encouraging development. This result was firstly shown by the observations of the distant SNe Ia
\cite{L1Riess,L2Perlmutter}, which can be seen as a standard candle \cite{DPhillips,DHamuy}. Afterwards, the CMB
measurement by Wilkinson Microwave Anisotropy Probe (WMAP) \cite{L3Spergel} and the large scale structure
survey by Sloan Digital Sky Survey (SDSS) \cite{L4Tegmark,L5Tegmark} support the same result as the SNe Ia
presented. To explain the acceleration of the universe, many cosmological models were introduced,
including the Quintessence \cite{L6Caldwell}, the brane world \cite{L7Deffayet}, the Chaplygin Gas \cite{L8Alcaniz} and the
holographic dark energy models \cite{L9Ke} and so on. The most popular model is referred as $\Lambda$CDM
cosmology composed of cold pressureless dark matter with the equation of state (EOS) $w=p/\rho$ and
Einstein's cosmological constant $\Lambda$ which is the most economic and the oldest form of dark
energy with $w=-1$ \cite{Peebles}. This model provides a reasonably good fit to most current
cosmological data \cite{RatraV,Frieman}. Additionally, one also consider another cosmological model--XCDM
parametrization which is useful in describing the time-varying dark energy models. In this model,
the dark energy is assumed to be a perfect fluid with the equation of state (EOS)
$\omega=p_{x}/\rho_{x}$, where $\omega$ is a number less than $-1/3$ \cite{Samushia2}. %$\omega<-1/3$.
In addition, the $f(R)$ gravity models are also constrained using the statistical lens sample from Sloan
Digital Sky Survey Quasar Lens Search Data Release 3 (SQLS DR3) \cite{YangFr}.

%Given a cosmological model, it is possible and convenient to compute quantities such as the Hubble parameter $H(z)$,
%which is a function of redshift $z$.
With the perfect observational data, one can compare the observational results
with theoretical predictions of different models and determine which model is better \cite{Samushia2,Samushia1}. Besides, another important
task of cosmology is to constrain the cosmological parameters of various cosmological models using the
redshift-dependent quantities, for example the luminosity distance to a particular class of objects such
as SNe Ia and Gamma-Ray Bursts (GRBs) \cite{L10Nesseris}.
X-ray gas mass fraction of galaxy clusters is also very popular \cite{L14Zhu}.
Recently, the size of the BAO peak detected in the large-scale
correlation function of luminous red galaxies from SDSS \cite{L12Eisentein} and the CMB
shift parameter $\mathcal {R}$ obtained from acoustic oscillations in the CMB temperature anisotropy power spectrum \cite{Hinshaw,Komatsu}
are also widely used to constrain cosmological models.
%The constraints using different
%data can provide some different results but more consistent with each other, and the combinations of varieties of data
%can also make the constraints tighter \cite{L15Li,L16Zhao,L17Xia}. Which one among the varieties of data plays a main role in
%the constraints is also a very important question.

Recently, one method based on the observational Hubble parameter $H(z)$ data as a function of redshift $z$,
which are related to the differential ages of the oldest galaxies has been used to test cosmological models
 \cite{L18Yi,Linhui,L19Samushia,L20Wei,L21Zhang,L22Wei,L23Zhang,L24Dantas,L25Zhang,L26Wu,L27Wei}.
  Furthermore, the new observations have given more OHD data  \cite{DalyH,CampoH}. These new released
data may improve the constraints of cosmological parameters evidently. Except that, the latest measurements
of the radial baryon acoustic oscillation (RBAO) were discussed deeply \cite{Samushia2,Gata}.
%But the RBAO measurements alone can not tightly constrain the cosmological parameters.
Lin et al's work \cite{Linhui} has shown that the constraints using different
data can provide some different results but more consistent with each other. And the combinations of varieties of data
can also make the constraints tighter \cite{L15Li,L16Zhao,L17Xia}. Following this direction, we intend to further examine the role
OHD played in constraining the cosmological parameters by using RBAO.
%and the application of the new data has
%given a helpful constraint.
%The motivation of this paper is to examine the role of OHD and SNe Ia in the constraint on cosmological parameters.
 We compare the constraints on the $\Lambda$CDM cosmology and XCDM cosmology using OHD
  and SNe Ia data combined with RBAO and CMB data. We find that the OHD plays the same role as SNe Ia for joint cosmological constraints.
%Depend on the identic of the constraints listed above, we give our discussion.

Our paper is organized as follows. In Sec.\ref{sec:data} we describe the observational data we used in this paper. In Sec.\ref{sec:model}
we present the dark energy models. In Sec.\ref{sec:constraint}, we show the constraints. %And in Sec.\ref{sec:discussion},
And finally, we give our conclusion.

\section{Observational data} \label{sec:data}

\subsection{SNe Ia data}

The luminosity distance of Type Ia supernova(SNe Ia), $d_{L}$, can be estimated by the relation

\begin{equation}
m=M+5\log d_{L}+25,
\end{equation}
where $m$ is the K-corrected observed apparent magnitude and $M$ the absolute magnitude of SNe Ia.
The luminosity distance depends on the content and geometry of the Universe in a
Friedmann-Robertson-Walker (FRW) cosmology

\begin{equation}
d_{L}=\frac{c(1+z)}{H_{0}\sqrt{|\Omega_{k}|}}{\rm sinn}[\sqrt{|\Omega_{k}|}\mathcal {F}(z)],
\end{equation}
where $sinn(x)$ is $sinh(x)$ for $\Omega_{k}>0$, $x$ for $\Omega_{k}=0$, and $sin(x)$ for $\Omega_{k}<0$
respectively. The function $\mathcal {F}(z)$ is defined as $\mathcal {F}(z)=\int^{z}_{0}dz/E(z)$ with
$E(z)=H(z)/H_{0}$. $E(z)$ is the expansion rate that has different forms in different cosmological
models. $H_{0}$ is the Hubble constant and $\Omega_{m}$, $\Omega_{\Lambda}$, $\Omega_{k}$ are the
matter, cosmological constant and curvature density parameters respectively. The distance modulus is

\begin{equation}
\mu_{z}=5{\rm log}\frac{d_{L}}{10pc}=42.39+5{\rm log}{\frac{1+z}{h\sqrt{|\Omega_k}|}{\rm sinn}[\sqrt{|\Omega_k}|\mathcal
{F}(z)]},
\end{equation}
where $h=H_{0}/{\rm 100km s^{-1} Mpc^{-1}}$. We use the observational SNe Ia data \cite{L28Riess} %which contains 185 data
with redshift spans from about 0.01 to 1.75.

\subsection{The Observational $H(z)$ data}

In addition, the measurement of the Hubble parameter $H(z)$ is increasingly becoming
important in cosmological constraints, and it can be derived from the
derivation of redshift $z$ with respect to the cosmic time $t$ \cite{L33Jimenez}

\begin{equation}
H(z)=-\frac{1}{1+z}\frac{dz}{dt},
\end{equation}
which provides a direct measurement for $H(z)$ through a determination of $dz/dt$. Jimenez et al
demonstrated the feasibility of the method by applying it to a $z\sim0$ sample\cite{L34Jimenez}. With the
availability of new galaxy surveys, it becomes possible to determine $H(z)$ at $z>0$. By using the
different ages of passively evolving galaxies determined from the Gemini Deep Deep Survey (GDDS) \cite{L35Abraham}
and archival data \cite{L36Treu,L37Treu,L38Nolan,L39Nolan}, Simon et al. derived a set of observational $H(z)$ data
\cite{L19Samushia,L34Jimenez,L40Simon}. The detailed estimation
method can be found in the work \cite{L40Simon}. As $z$ has a relatively wide range, $0.1<z<1.8$, these data are
expected to provide a more full-scale description of the dynamical evolution of our universe. The
application of OHD to cosmology can be referred to \cite{L18Yi,L19Samushia,L20Wei,Linhui,L40Simon}.

\subsection{The CMB data}

%The CMB shift parameter $\mathcal {R}$ is one of the most model-independent
%parameter which can be derived from CMB data and not dependent on $H_{0}$ \cite{L30Bond,L31Odman}
The CMB shift parameter $\mathcal {R}$ is arguably one of the most model-independent parameters
among those which can be inferred from CMB data, provided that the dark energy density parameter is negligible at
recombination, and it does not depend on $H_{0}$ \cite{L30Bond,L31Odman}. It is directly proportional to the
ratio of the angular diameter distance to the decoupling epoch divided by the Hubble horizon size at the decoupling epoch.
That is,

\begin{equation}
\mathcal {R}=\frac{\sqrt{\Omega_{m}}}{\sqrt{|\Omega_{k}|}}{\rm sinn}[\sqrt{|\Omega_{k}|}\mathcal
{F}(z_{s})],
\end{equation}
where $z_{s}=1089$ is the redshift of recombination. The value of $\mathcal {R}$ obtained from acoustic oscillations
in the CMB temperature anisotropy power spectrum is $\mathcal {R}=1.715\pm0.021$ \cite{Hinshaw,Komatsu}.

\subsection{The RBAO data}

The measurement of the large-scale structure baryon acoustic
oscillation (BAO) peak length scale has been found efficient to constrain cosmological parameters \cite{SamushiaN,Ishida,Lazkoz,Santos}.
Gazta{$\rm \tilde{n}$}aga recently used SDSS data to measure the radial baryon acoustic scale in two
redshift ranges $z\sim0.15-0.30$. The radial baryon acoustic scale is independent from the previous BAO measurement which
was either averaged over all direction or just in the transverse direction.
The data used was listed in their Table.I in \cite{Gata}.
Theoretically the radial BAO peak scale is given by

\begin{equation}
\Delta z=H(z)r_{s}/c,
\end{equation}
where $H(z)$ is the Hubble parameter at redshift $z$, $r_{s}$ is the sound horizon at the time of
recombination, and $c$ is the speed of light respectively.
%To compute $r_{s}$, Gazta{$\rm \tilde{n}$}aga use two
%different methods. We choose the first one, which is to use the WMAP measured ratio $l_{s}$ between the distance to the last-scattering surface.
%and $r_{s}$ measured by CMB anisotropy experiments.
%The other option can give the similar drawback \cite{Samushia2}.
$r_{s}$ can be computed as \cite{Samushia2}

\begin{equation}
r_{s}=\frac{\pi(1+z)d_{A}(z_{s})}{l_{s}},
\end{equation}
where $z_{s}$ is the redshift of the last-scattering surface. We adopt the WMAP 5-year recommended
values $l_{s}=302.14\pm0.87, z_{s}=1090.5\pm1.0$. While $d_{A}$ is the angular diameter
distance, it can be easily computed in a given cosmological model.
%=================================================

\section{Cosmology models}  \label{sec:model}

We apply the data listed in Sec.II with the predictions of two cosmological models including dark energy. The models we
consider are standard $\Lambda$CDM and XCDM parametrization of the dark energy's equation of state.
As mentioned above, the difference of the %expansion rate $E(z)$ between these dark energy models is
two models is existed in the expansion rate

\begin{equation}
\begin{array}{llll}
E(z)=\sqrt{\Omega_{m}(1+z)^{3}+\Omega_{\Lambda}+\Omega_{k}(1+z)^2} \hspace{15pt}     (\Lambda CDM)  \\
\\
E(z)=\sqrt{\Omega_{m}(1+z)^{3}+(1-\Omega_{m})(1+z)^{3(1+\omega)}}   \hspace{8pt}      (XCDM)
\end{array}
\end{equation}
%\begin{equation}
%E(z)=\sqrt{\Omega_{m}(1+z)^{3}+(1-\Omega_{m})(1+z)^{3(1+\omega)}}   \hspace{8pt}      (XCDM)
%\end{equation}
In both of the two models the background evolution is described by two parameters. One is the
nonrelativistic matter fractional energy density parameter $\Omega_{m}$ %which works the same in both models
 and the other one is a parameter
that characterizes the dark energy. For the $\Lambda$CDM model, the parameter is the cosmological constant
fractional energy density parameter $\Omega_{\Lambda}$. In the XCDM parametrization, it is the
equation of state parameter $\omega$. In this paper, we assume that the XCDM model is the
spatially-flat while in the $\Lambda$CDM case, the spatial curvature is allowed to vary, with the
space curvature fractional energy density parameter $\Omega_{k}=1-\Omega_{m}-\Omega_{\Lambda}$.
%So the angular diameter distance can be computed as

%\begin{equation}
%\begin{array}{lllll}
%d_{A}(z)=\frac{1}{\sqrt{\Omega_{k}}H_{0}(1+z)}{\sinh}[\sqrt{\Omega_{k}}\int^{z}_{0}\frac{dz'}{E(z')}]
% \hspace{8pt} (\Omega_{k}>0)  \\
%\\
%d_{A}(z)=\frac{1}{H_{0}(1+z)}\int^{z}_{0}\frac{dz'}{E(z')}     \hspace{25pt}     (\Omega_{k}=0)  \\
%\\
%d_{A}(z)=\frac{1}{\sqrt{-\Omega_{k}}H_{0}(1+z)}{\sin}[\sqrt{-\Omega_{k}}\int^{z}_{0}\frac{dz'}{E(z')}]
%  \hspace{8pt}  (\Omega_{k}<0)
% \end{array}
%\end{equation}

%\begin{equation}
%d_{A}(z)=\frac{1}{H_{0}(1+z)}\int^{z}_{0}\frac{dz'}{E(z')}     \hspace{25pt}     (\Omega_{k}=0)
%\end{equation}

%\begin{equation}
%d_{A}(z)=\frac{1}{\sqrt{-\Omega_{k}}H_{0}(1+z)}{\sin}[\sqrt{-\Omega_{k}}\int^{z}_{0}\frac{dz'}{E(z')}]
%  \hspace{8pt}  (\Omega_{k}<0)
%\end{equation}

\section{Cosmological constraints}  \label{sec:constraint}

\begin{figure}
%\vspace{.2in}
\centerline{\psfig{figure=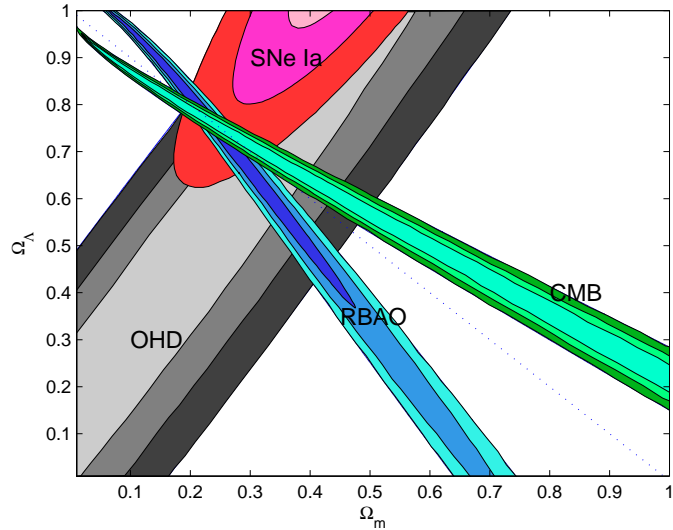,width=4truein,height=3truein}
\hskip 0.0in} \caption{Confidence regions in the $\Omega_{m}-\Omega_{\Lambda}$ plane for the data used alone for $\Lambda$CDM model.
For each kind of data, with a prior of $h$,
 the confidence regions at 68.3\%, 95.4\% and 99.7\% levels from inner to outer are presented respectively.
 The dotted line demarcates spatially-flat {\rm $\Lambda$}CDM models.} \label{fig:LCDMfull}
\end{figure}

The purpose of this paper is to examine the role of OHD and SNe Ia data in the constraints on
cosmological parameters with different cosmological models. The likelihood for the cosmological
parameters can be determined from a $\chi^{2}(h,\Omega_{m},\Omega_{\Lambda})$ statistics. For RBAO in $\Lambda$CDM as an example,
\begin{equation}
\chi^{2}(\Omega_{m},\Omega_{\Lambda},h)=\sum_{i=1,2}(\Delta z_{th,i}(\Omega_{m},\Omega_{\Lambda},h)-\Delta z_{obs,i})^{2}/\sigma_{\Delta z,i}^{2},
\end{equation}
to get the
constraint results of two parameters $\Omega_{m}$ and $\Omega_{\Lambda}$, we marginalize the likelihood functions over $h$ by integrating
the probability density $P\propto e^{\chi^{2}/2}$. Thus we can obtain the best fitting results and
the confidence regions in the $\Omega_{m}-\Omega_{\Lambda}$ plane.
%At present, the RBAO scale is measured only in two redshift ranges and does not provide very efficient constraints. However,
%it is independent from the Hubble parameter as the CMB data.
%These two data can be the basis of the comparison of the Hubble parameter and SNe Ia data.
 In the calculation, we assume a prior of $h=0.705\pm0.013$ as WMAP5 suggested, and this method
can improve the constraints greatly \cite{Linhui}.
Fig.\ref{fig:LCDMfull} shows the constraints of each data without any combination for $\Lambda$CDM model. It can be easily seen that
the confidence regions for each data are almost different. And the best fitting results they give are also different.
Both RBAO and CMB data prefer a nearly flat universe.
%Their best fitting values reflect that the space curvature fractional energy density parameter $\Omega_{k}$ approaches zero.
%Table I shows the fitting results for the $\Lambda$CDM model with a prior of $h$.

%===================figure1=======================
%=================================================

%======================figure2====================
\begin{figure*}
%\vspace{.2in}
\centerline{\psfig{figure=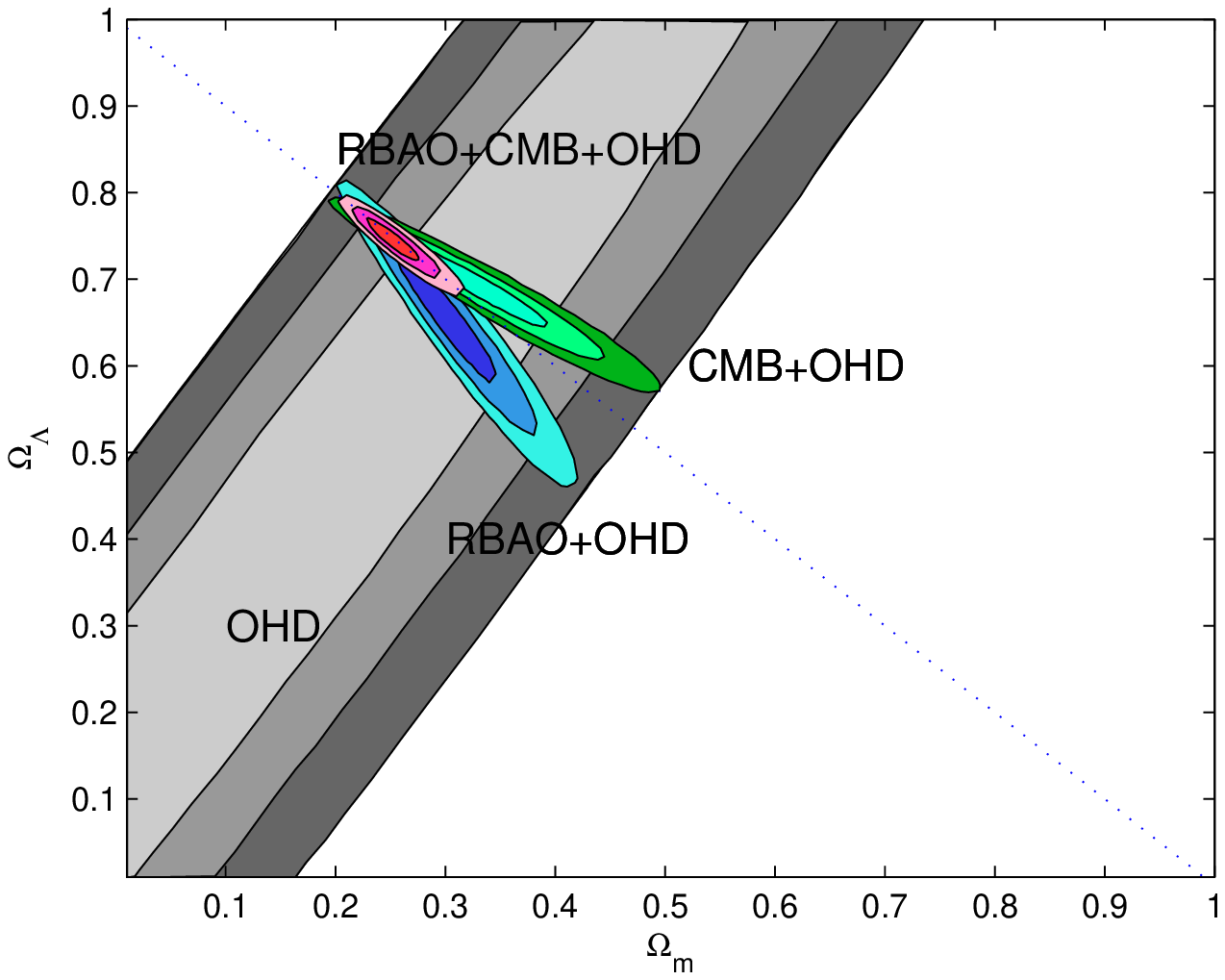,width=4truein,height=3truein}
\psfig{figure=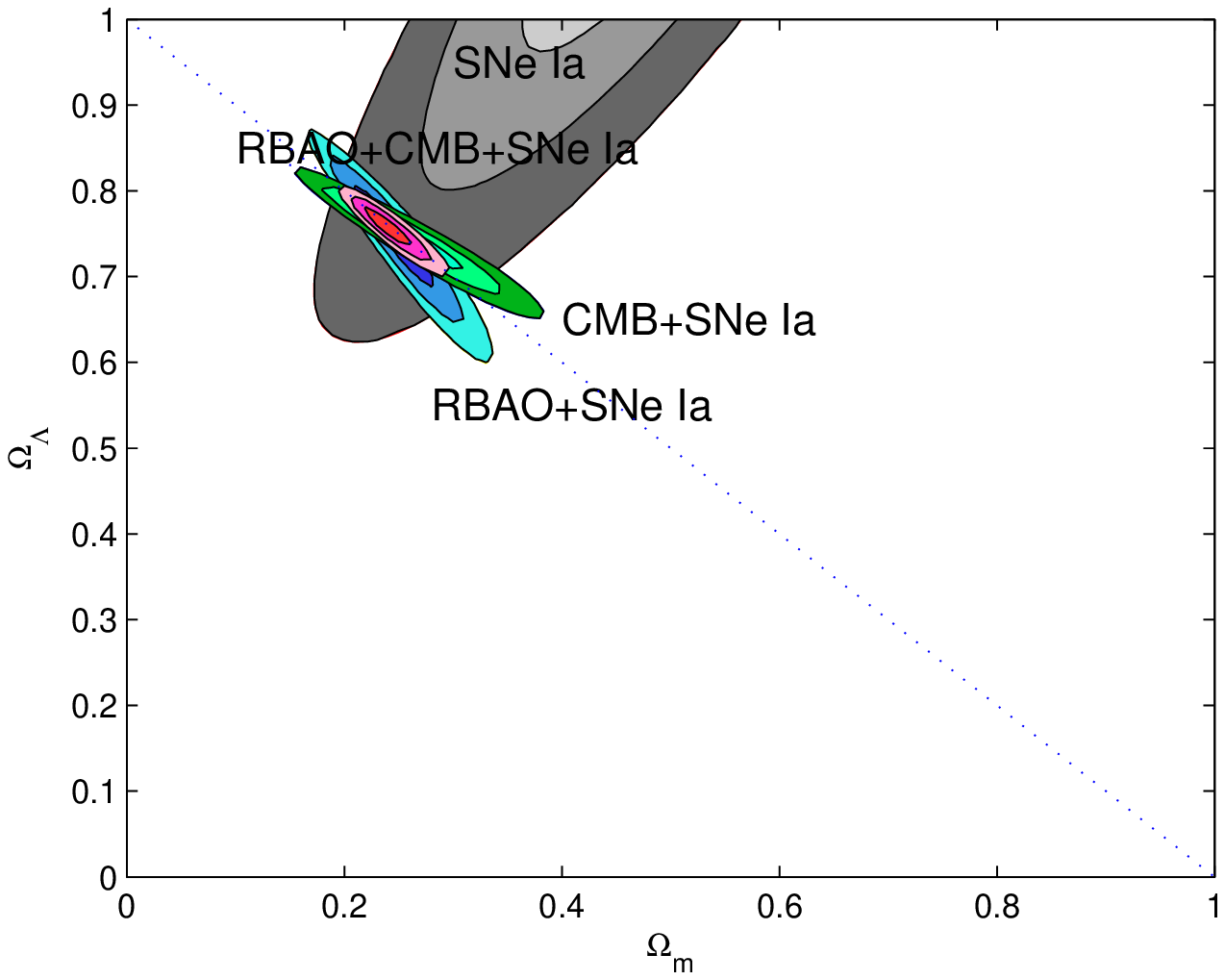,width=4truein,height=3truein}
\hskip 0.1in}
\caption{The confidence regions of the combined constraints of $\Lambda$CDM model.
$Left$ $panel$: The OHD data combined with RBAO and CMB respectively. And the smallest one correspond to the constraint of the data of
RBAO+CMB+OHD. $Right$ $panel$: The SNe Ia data combined with RBAO and CMB respectively.
The smallest one indicates the constraint of the data of RBAO+CMB+SNe Ia.
The dotted line demarcates spatially-flat {\rm $\Lambda$}CDM models.}\label{fig:LCDM}
\end{figure*}
%=================================================

%======================figure3====================
\begin{figure*}
%\vspace{.2in}
\centerline{\psfig{figure=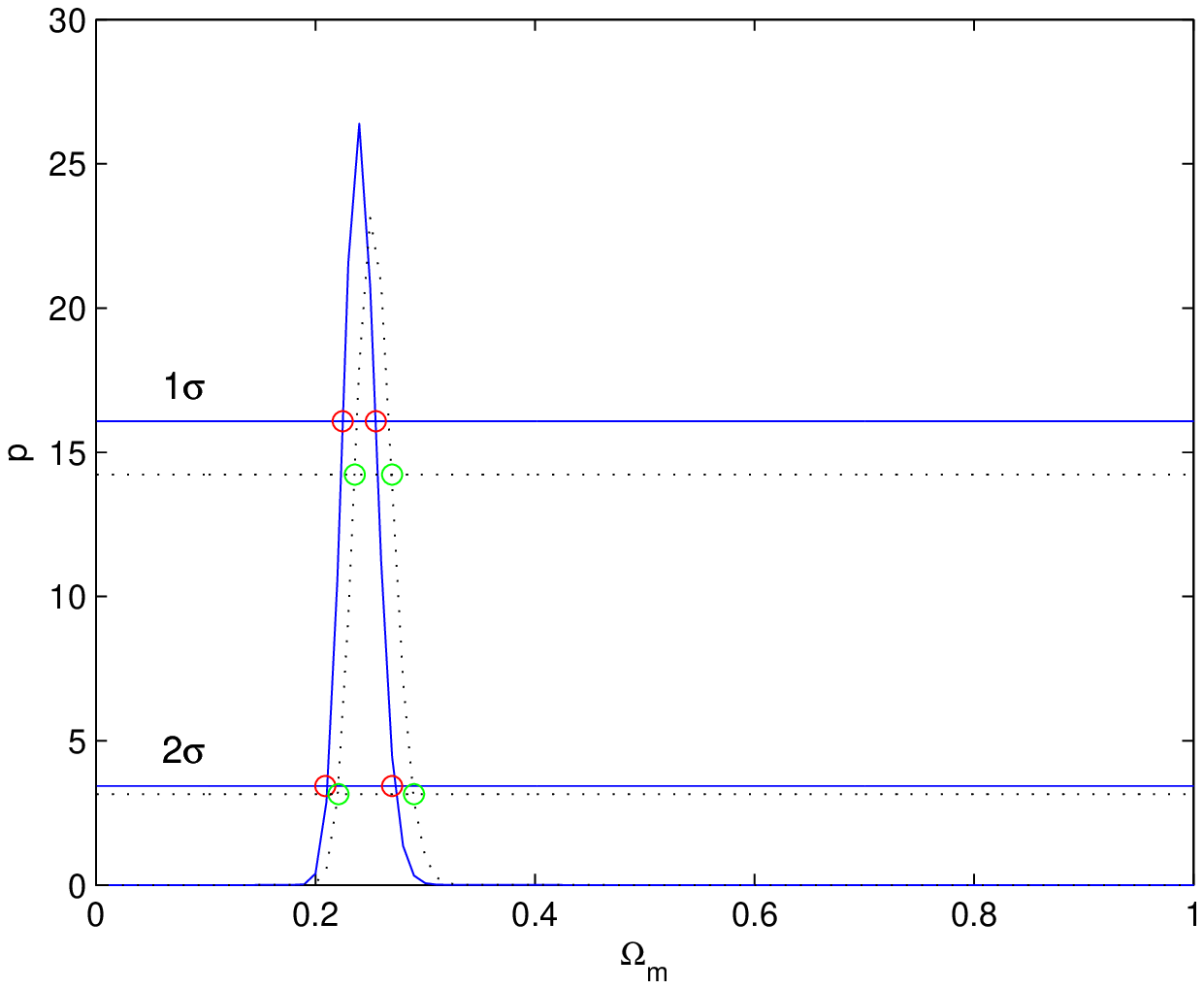,width=4truein,height=3truein}
\psfig{figure=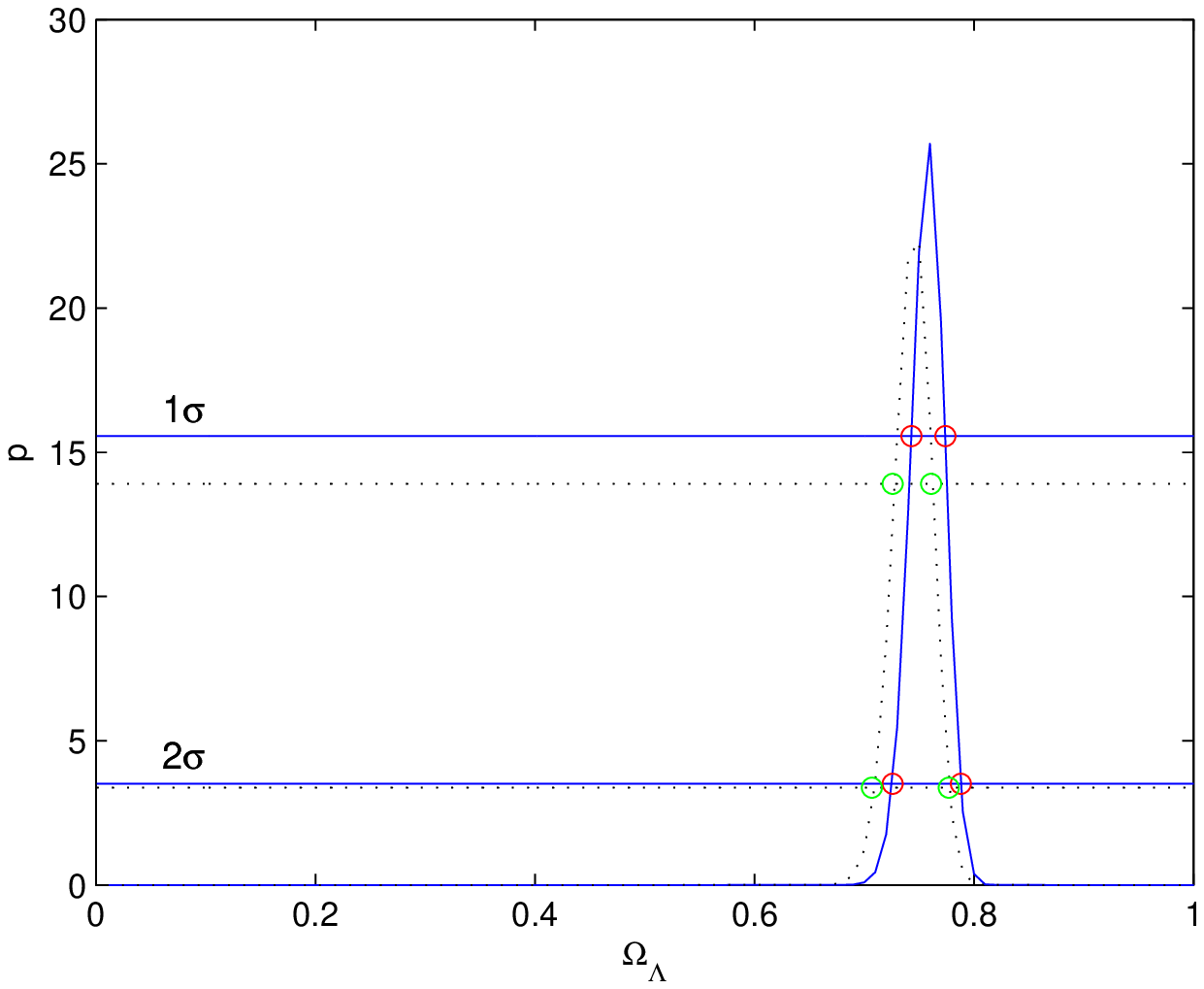,width=4truein,height=3truein}
\hskip 0.1in}
\caption{The one-dimensional probability distribution function(PDF)$p$ with the data of RBAO+CMB+OHD (dotted line)
and RBAO+CMB+SNe Ia (solid line) for selections of parameters $\Omega_{m}$ and $\Omega_{\Lambda}$ with a prior of $h$ for $\Lambda$CDM model.
The 1 and 2 $\sigma$ confidence levels are also labeled.
}\label{fig:LCDMOHD}
\end{figure*}
%=================================================

%%======================figure4====================
%\begin{figure*}
%\vspace{.2in}
%\centerline{\psfig{figure=LCDMSNePDFOM.eps,width=4truein,height=3truein}
%\psfig{figure=LCDMHUBBLEPDFOL.eps,width=4truein,height=3truein}
%\psfig{figure=LCDMSNePDFOM.eps,width=4truein,height=3truein}
%\psfig{figure=LCDMSNePDFOL.eps,width=4truein,height=3truein}
%\hskip 0.1in}
%\caption{The one-dimensional probability distribution function(PDF)$p$ with the data of RBAO+CMB+SNe Ia
%for selections of parameters $\Omega_{m}$ and $\Omega_{\Lambda}$ with a prior of $h$ for $\Lambda$CDM model.
%The 1 and 2 $\sigma$ confidence levels are also labeled.
%}\label{fig:LCDMSNe}
%\end{figure*}
%=================================================

%===================figure4=======================
\begin{figure}
%\vspace{.2in}
\centerline{\psfig{figure=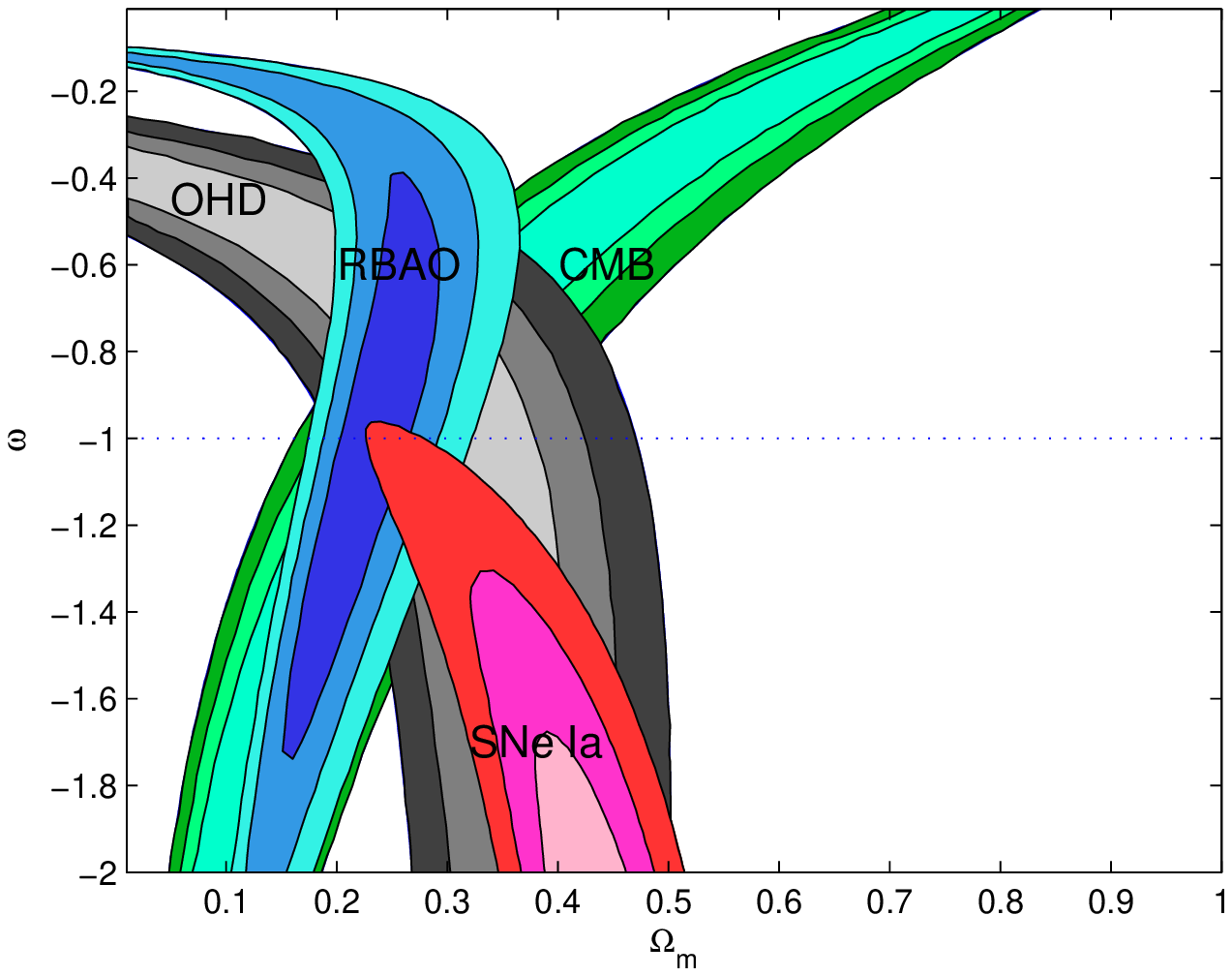,width=4truein,height=3truein}
\hskip 0.0in}
\caption{Same as Fig.\ref{fig:LCDMfull}, but for XCDM.} \label{fig:XCDMfull}
\end{figure}
%=================================================

%======================figure5====================
\begin{figure*}
%\vspace{.2in}
\centerline{\psfig{figure=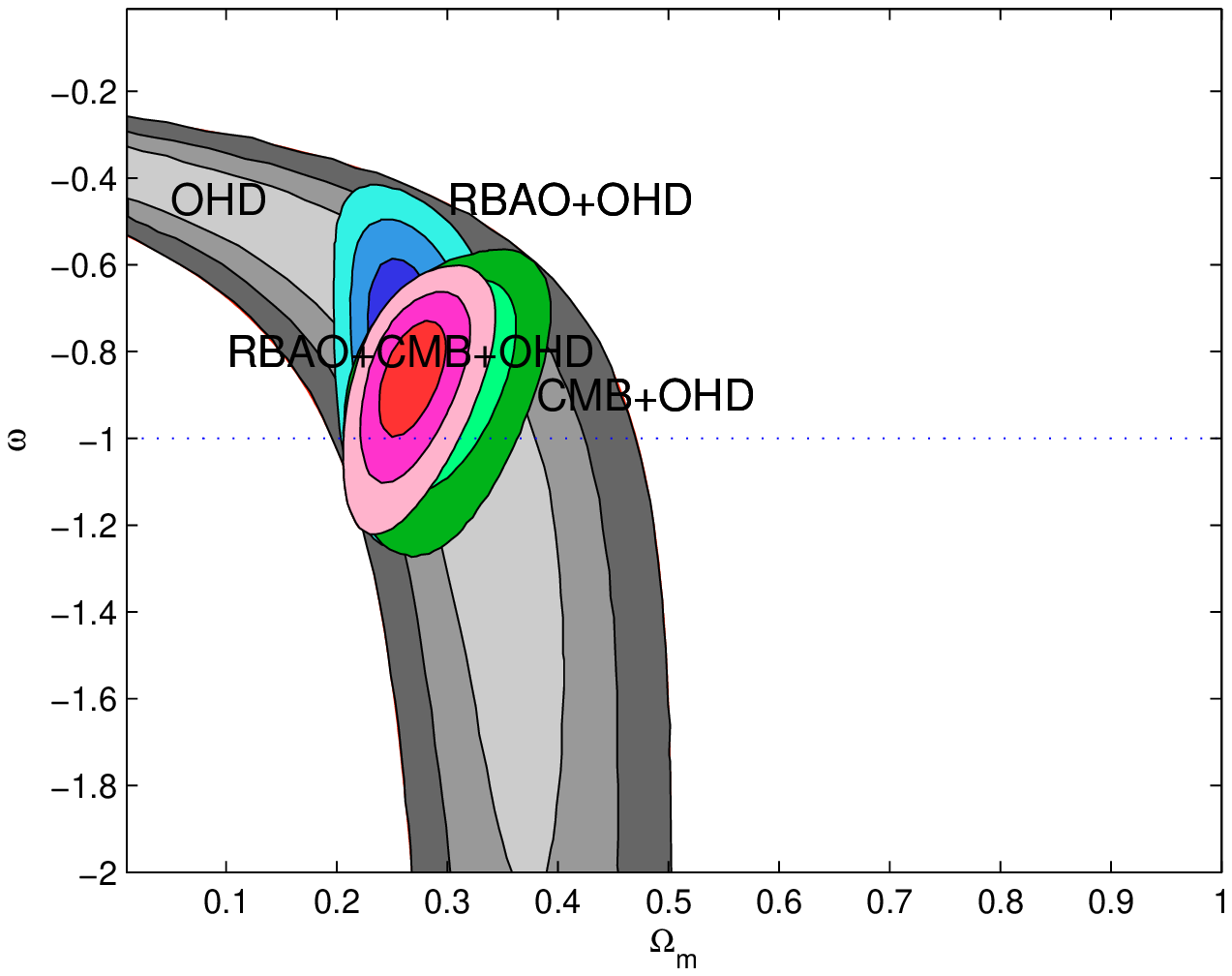,width=4truein,height=3truein}
\psfig{figure=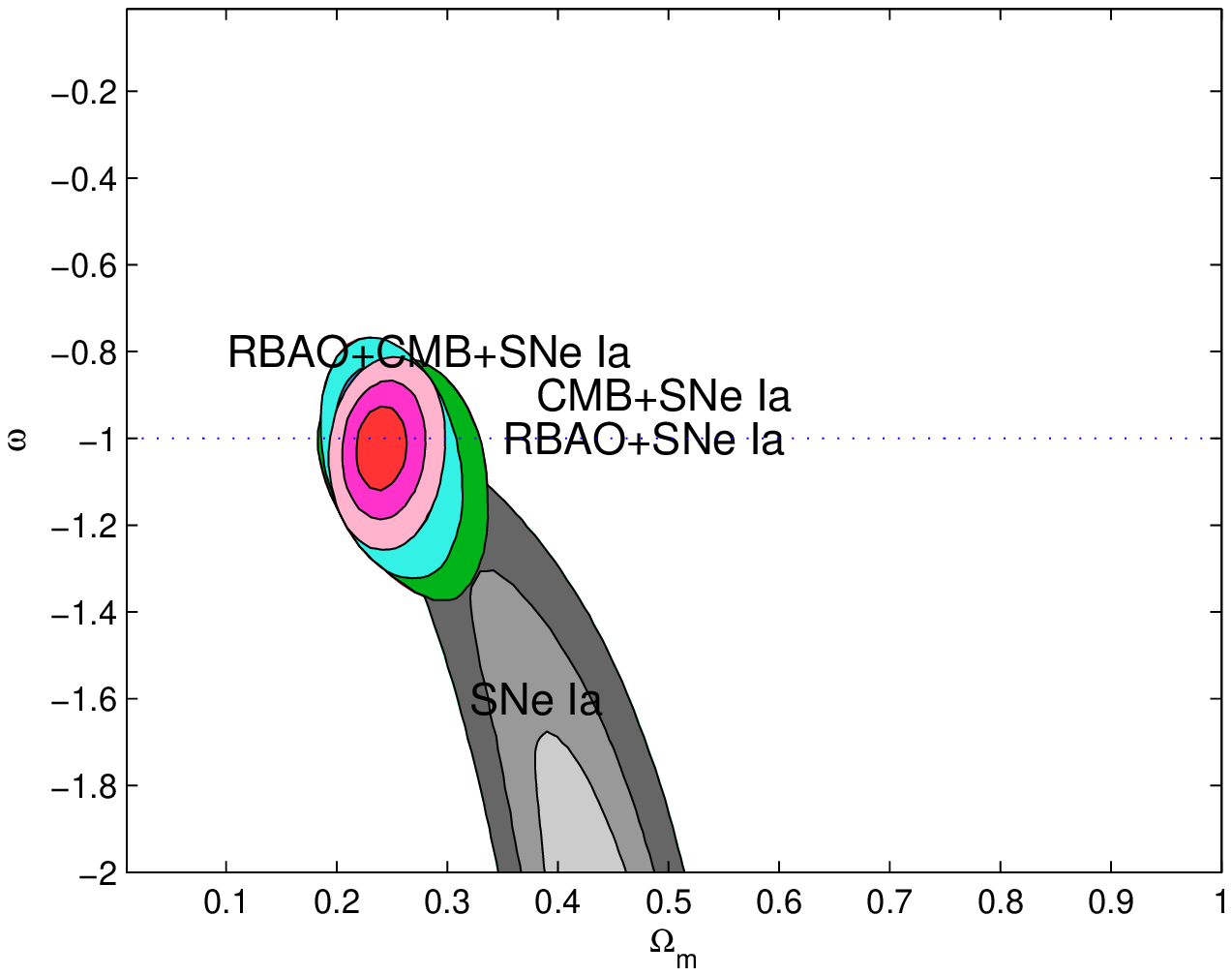,width=4truein,height=3truein}
\hskip 0.1in}
\caption{Same as Fig.\ref{fig:LCDM}, but for XCDM.}\label{fig:XCDM}
\end{figure*}
%=================================================

%======================figure6====================
\begin{figure*}
%\vspace{.2in}
\centerline{\psfig{figure=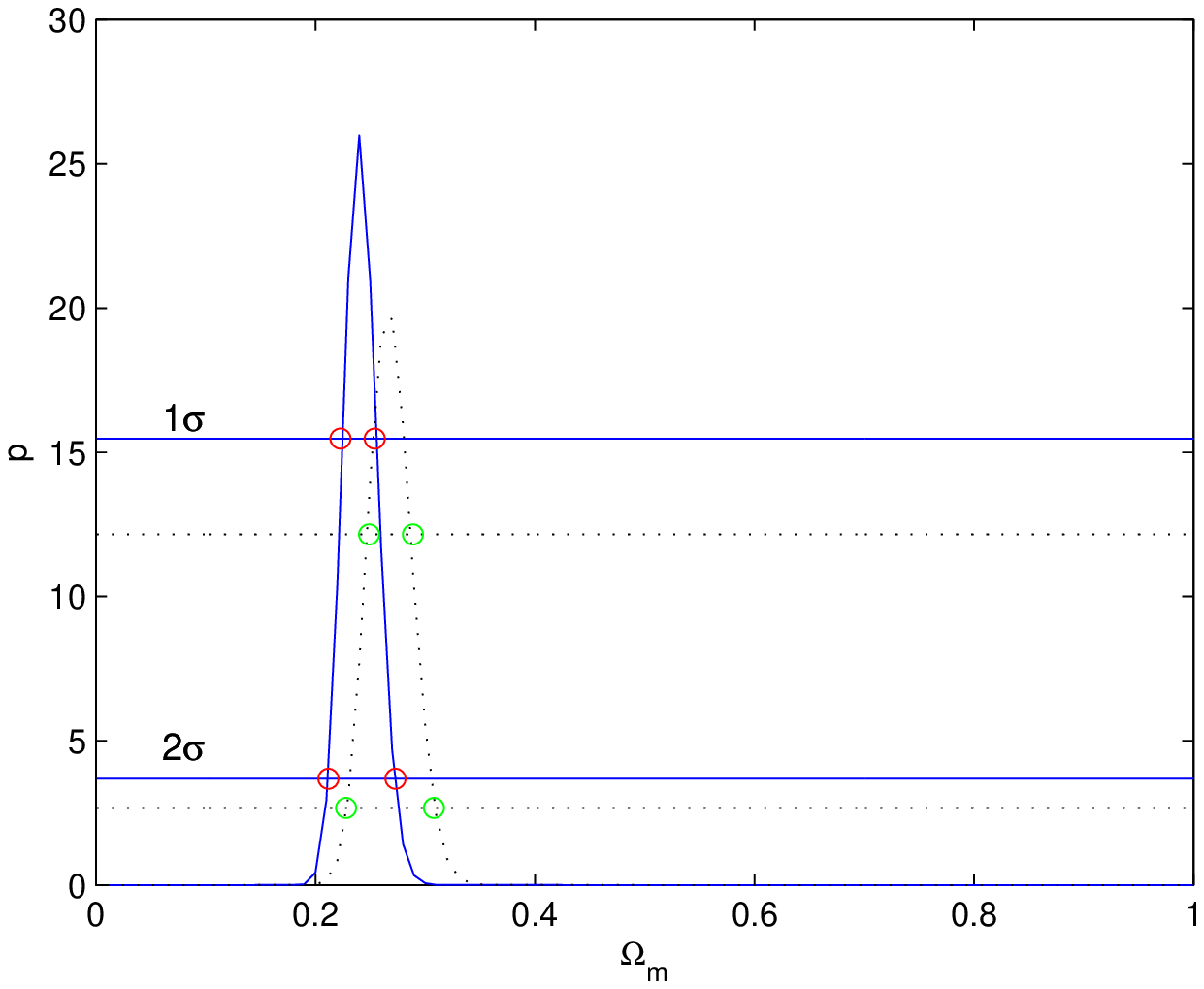,width=4truein,height=3truein}
\psfig{figure=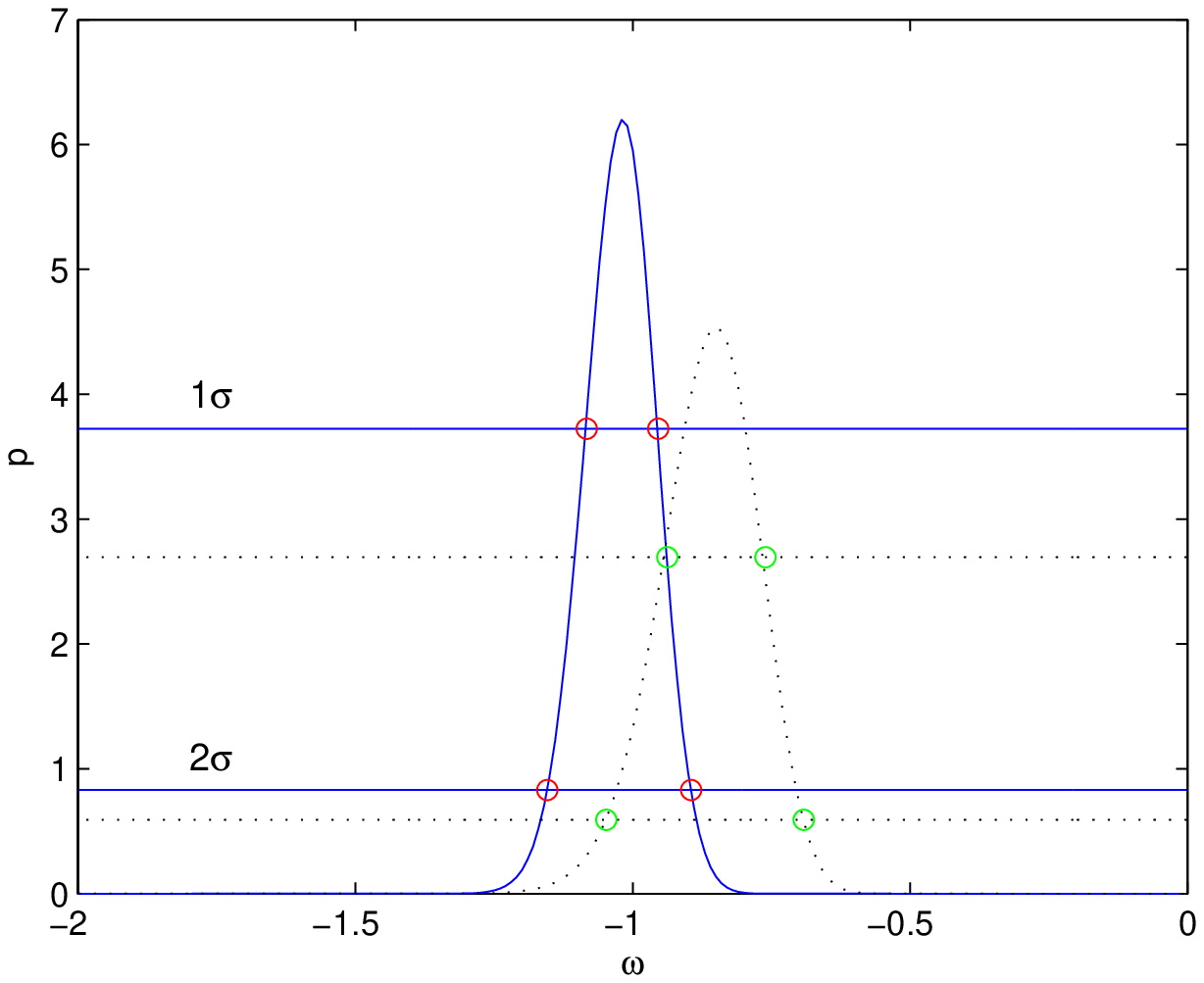,width=4truein,height=3truein}
\hskip 0.1in}
\caption{Same as Fig.\ref{fig:LCDMOHD}, but for XCDM.}\label{fig:XCDMOHD}
\end{figure*}
%=================================================

%%======================figure7====================
%\begin{figure*}
%%\vspace{.2in}
%\centerline{\psfig{figure=XCDMSNePDFOM.eps,width=4truein,height=3truein}
%\psfig{figure=XCDMSNePDFOL.eps,width=4truein,height=3truein}
%\hskip 0.1in}
%\caption{Same as Fig.\ref{fig:LCDMSNe}, but for XCDM.}\label{fig:XCDMSNe}
%\end{figure*}
%%=================================================

In order to compare the contribution of OHD and SNe Ia data in constraining the cosmological parameters clearly, it is effective
to combine the different data together. Fig.\ref{fig:LCDM} presents the combined constraints of OHD and SNe Ia data with RBAO and CMB respectively.
It is shown from the Table.\ref{tab:lcdm} that there are slight differences on the best fitting values of $\Omega_{m}$ and $\Omega_{\Lambda}$.
However, the consistency of the results which is more important
%and the changing tendency of the combined constraints
indicates that the OHD and SNe Ia data give almost the same contribution in constraining cosmological parameters.
%especially for the RBAO+CMB+OHD and RBAO+CMB+SNe Ia although some slight discrepancy on the confidence regions of 68.3\%, 95.4\% and 99.7\% levels
%between them are revealed in the figure.
We also calculate the one-dimensional probability distribution function(PDF)$p$ for selections
of parameters $\Omega_{m}$ and $\Omega_{\Lambda}$ with a prior of $h$. Fig.\ref{fig:LCDMOHD}
presents the PDF of $\Omega_{m}$ and $\Omega_{\Lambda}$ for RBAO+CMB+OHD and RBAO+CMB+SNe Ia respectively.
The 1 and 2$\sigma$ confidence levels are also shown.
%Besides, the PDF should be
%unitary, which means that the area under the PDF curve should equal 1. However, it is important to get the relative PDF rather than the
%absolute one. So one can set the maximum of the PDF equal 1. It will not influence the analyze of the constraints.
It is easy to see that the most probable value of the two results are roughly consistent with each other.

From the constraints of combined data and the one-dimensional probability distribution function, we can see that some slight
discrepancy are shown between the constraints of OHD and SNe Ia combined with other data. However, both the constraints are almost the same restrictive.
And their results prefer a nearly flat universe.
Applying the data we used above to the XCDM model,
% we can also get a series of the calculating results.
first we get Fig.\ref{fig:XCDMfull} that shows the constraints of the alone data used in Fig.\ref{fig:LCDMfull}.
While Fig.\ref{fig:XCDM} shows the combined constraints as Fig.\ref{fig:LCDM}.
%The OHD and SNe Ia data are combined with RBAO and CMB respectively.
It is clear that the constraints of RBAO+CMB+OHD and RBAO+CMB+SNe Ia are both restrictive at the confidence level of 68.3\%. The best-fit results of
these constraints are listed in Table.\ref{tab:xcdm}. In order to examine if the OHD and SNe Ia data play the same role in constraining the
cosmological parameters as in $\Lambda$CDM model, the one-dimensional probability distribution function(PDF) is also calculated.
The PDF curves are plotted in Fig.\ref{fig:XCDMOHD}. From the results listed above, we can
see that the constraints of the parameter $\Omega_{m}$ using the two data combinations are consistent with each other. The main
discrepancy is in constraining $\omega$. The 1$\sigma$ confidence region of $\omega$ achieved from RBAO+CMB+OHD is $\omega=-0.84\pm0.14$,
 while RBAO+CMB+SNe Ia suggests $\omega=-1.02\pm0.10$.
 %The constraint of RBAO+CMB+SNe Ia is consistent with the result of data of Gamma-ray bursts\cite{L11Dai}.
 The reason that causes the difference is that the amount of the OHD is so few.
%However, the results of OHD and SNe Ia both prefer a universe dominated by dark energy.
With more data achieved in the future, many deficiencies will be improved.
%==================== table 1 ====================
\begin{table}
\begin{center}
\begin{tabular}{ccc}
\hline\hline \
parameters  & $\Omega_{m}$ & $\Omega_{\Lambda}$ \\
RBAO+OHD & $0.30\pm0.04$ & $0.66\pm0.07$    \\
CMB+OHD & $0.32\pm0.06$ & $0.70\pm0.04$  \\
RBAO+CMB+OHD & $0.25\pm0.02$ & $0.75\pm0.03$  \\
\\
RBAO+SNe Ia & $0.25\pm0.03$ & $0.74\pm0.06$ \\
CMB+SNe Ia & $0.25\pm0.04$ & $0.75\pm0.03$ \\
RBAO+CMB+SNe Ia & $0.24\pm0.02$ & $0.76\pm0.02$\\
\hline\hline
\end{tabular}
\caption{The best-fit results of the $\Lambda$CDM model with a prior of $h$.} \label{tab:lcdm}
\end{center}
\end{table}
%=================================================

%==================== table 2 ====================
\begin{table}
\begin{center}
\begin{tabular}{cccc}
\hline\hline \
parameters  & $\Omega_{m}$ & $\omega$ \\
RBAO+OHD & $0.25\pm0.03$ & $-0.75\pm0.18$    \\
CMB+OHD & $0.29\pm0.04$ & $-0.84\pm0.15$  \\
RBAO+CMB+OHD & $0.27\pm0.03$ & $-0.84\pm0.14$  \\
\\
RBAO+SNe Ia & $0.24\pm0.03$ & $-1.02\pm0.11$ \\
CMB+SNe Ia & $0.25\pm0.03$ & $-1.03\pm0.12$ \\
RBAO+CMB+SNe Ia & $0.24\pm0.02$ & $-1.02\pm0.10$\\
\hline\hline
\end{tabular}
\caption{The best-fit results of the XCDM model with the a prior of $h$.} \label{tab:xcdm}
\end{center}
\end{table}
%================================================

\section{Discussions and Conclusions} \label{sec:discussion}
Recent observations have provided a lot of information to analyze the dynamical behavior of the universe. Most of them are based on distance
measurements, such as SNe Ia. And the present BAO peak scale data and CMB data are both
 sparse and can not provide a tight constraint on dark energy parameters. It is important to use other different probes to set
bonds on the cosmological parameters. Following this direction,
we used the observational $H(z)$ data from the differential ages
of the passively evolving galaxies to constrain the $\Lambda$CDM cosmology and XCDM cosmology, combining RBAO and CMB. In order to verify
the OHD data can give almost the same contribution in constraining the cosmological parameters as other widely used data,
we compared the SNe Ia data in the same way of calculation.

For the $\Lambda$CDM universe with a prior of $h$, the best-fit result of RBAO+CMB+OHD and RBAO+CMB+SNe Ia indicate a nearly flat universe.
The constraints of these two data combinations are both very tight and consistent with each other.
For the flat XCDM universe with the same prior, there exists some discrepancy in the constraints, especially for
the parameter $\omega$, however, the constraint results of $\Omega_{m}$ obtained from
the two data combination RBAO+CMB+OHD and RBAO+CMB+SNe Ia are almost the same.

From the above comparison and previous works \cite{L18Yi,L19Samushia},
we find that our results from the observational $H(z)$ data are believable and the computation results
are consistent with the results using the data of SNe Ia. So the observational $H(z)$ data can be seen as a
complementarity to other cosmological probes. With a large amount of OHD in a wider range of redshift $z$ in the future, we probably can
constrain the cosmological parameters using OHD instead or combined with other data.

\section{Acknowledgments}

%We are very grateful to the anonymous referee for valuable comments
%and suggestions that greatly improve this paper.
We are very grateful to the anonymous referee for his valuable comments and suggestions that greatly improve this paper.
Z.X.Z would like
to thank Hao-Ran Yu, Qiang-Yuan and Ze-Long Yi for their kindly help and very helpful suggestions and discussions.
This work was supported by the National Science Foundation of China (Grants No.10473002), the Ministry of Science and Technology
National Basic Science program (project 973) under grant No.2009CB24901, the Scientific Research Foundation of Beijing Normal University,
the Scientific Research Foundation for the Returned Overseas
Chinese Scholars, State Education Ministry.

%T.J.Z. would like to thank Hui-Lin for her useful comments.
%This work is done under the great help of Scientific Computing Center of BNU.
%This research was supported in part by the Project of Knowledge Innovation Program
%(PKIP) of Chinese Academy of Sciences, Grant No. KJCX2.YW.W10

\end{document}